\documentclass[journal]{IEEEtran}
\usepackage{url}
\usepackage[utf8]{inputenc}
\usepackage{xcolor}
\usepackage{amsmath}
\usepackage{amssymb}
\usepackage{microtype}

\usepackage[acronyms,nonumberlist,nopostdot,nomain,nogroupskip]{glossaries}
\usepackage{tablefootnote}
\usepackage{booktabs}
\usepackage{tabularx}
\usepackage{tikz}
\usepackage{pgfplots}
\pgfplotsset{compat=newest} 
\pgfplotsset{plot coordinates/math parser=false} 
\newlength\fheight
\newlength\fwidth
\usetikzlibrary{plotmarks,patterns,decorations.pathreplacing,backgrounds,calc,arrows,arrows.meta,spy,matrix}
\usepgfplotslibrary{patchplots,groupplots}
\usepackage{tikzscale}

\usepackage{multirow}
\usepackage{tkz-kiviat}
\usepackage[htt]{hyphenat}

\usepackage[font=scriptsize]{subcaption}
\usepackage[font=footnotesize]{caption}

\usepackage{mathtools}

\usepackage{dblfloatfix}    
\usepackage{colortbl}

\usepackage{makecell}
\usepackage{diagbox}
\usepackage{tikz-qtree}
\usetikzlibrary{trees} 
\newacronym{3gpp}{3GPP}{3rd Generation Partnership Project}
\newacronym{adc}{ADC}{Analog to Digital Converter}
\newacronym{5g}{5G}{5th generation}
\newacronym{aimd}{AIMD}{Additive Increase Multiplicative Decrease}
\newacronym{am}{AM}{Acknowledged Mode}
\newacronym{amc}{AMC}{Adaptive Modulation and Coding}
\newacronym{aqm}{AQM}{Active Queue Management}
\newacronym{awgn}{AGWN}{Additive White Gaussian Noise}
\newacronym{balia}{BALIA}{Balanced Link Adaptation}
\newacronym{bdp}{BDP}{Bandwidth-Delay Product}
\newacronym{bf}{BF}{beamforming}
\newacronym{cc}{CC}{Congestion Control}
\newacronym{cdf}{CDF}{Cumulative Distribution Function}
\newacronym{cn}{CN}{Core Network}
\newacronym{cqi}{CQI}{Channel Quality Information}
\newacronym{cp}{CP}{Control Plane}
\newacronym{csirs}{CSI-RS}{Channel State Information - Reference Signal}
\newacronym{dc}{DC}{Dual Connectivity}
\newacronym{rb}{RB}{Resource Block}
\newacronym{dce}{DCE}{Direct Code Execution}
\newacronym{dci}{DCI}{Downlink Control Information}
\newacronym{udp}{UDP}{User Datagram Protocol}
\newacronym{dl}{DL}{Downlink}
\newacronym{dmr}{DMR}{Deadline Miss Ratio}
\newacronym{dmrs}{DMRS}{DeModulation Reference Signal}
\newacronym{e2e}{E2E}{End-to-End}
\newacronym{si}{SI}{Study Item}
\newacronym{ecn}{ECN}{Explicit Congestion Notification}
\newacronym{edf}{EDF}{Earliest Deadline First}
\newacronym{enb}{eNB}{eNodeB}
\newacronym{epc}{EPC}{Evolved Packet Core}
\newacronym{es}{ES}{Edge Server}
\newacronym{cav}{CAV}{Connected and Autonomous Vehicle}
\newacronym{fdma}{FDMA}{Frequency Division Multiple Access}
\newacronym{fdd}{FDD}{Frequency Division Duplexing}
\newacronym{upa}{UPA}{Uniform Planar Array}
\newacronym[firstplural=Radio Access Technologies (RATs)]{rat}{RAT}{Radio Access Technology}
\newacronym[firstplural=Radio Access Technology (RTs)]{rt}{RT}{Radio Technology}
\newacronym{fs}{FS}{Fast Switching}
\newacronym{ftp}{FTP}{File Transfer Protocol}
\newacronym{gnb}{gNB}{Next Generation Node Base}
\newacronym{harq}{HARQ}{Hybrid Automatic Repeat reQuest}
\newacronym{hetnet}{HetNet}{Heterogeneous Network}
\newacronym{hh}{HH}{Hard Handover}
\newacronym{hol}{HOL}{Head-of-Line}
\newacronym{ia}{IA}{Initial Access}
\newacronym{imt}{IMT}{International Mobile Telecommunication}
\newacronym{iot}{IoT}{Internet of Things}
\newacronym{los}{LOS}{Line of Sight}
\newacronym{lte}{LTE}{Long Term Evolution}
\newacronym{m2m}{M2M}{Machine to Machine}
\newacronym{mac}{MAC}{Medium Access Control}
\newacronym{mc}{MC}{Multi-Connectivity}
\newacronym{mcs}{MCS}{Modulation and Coding Scheme}
\newacronym{mec}{MEC}{Mobile Edge Cloud}
\newacronym{mi}{MI}{Mutual Information}
\newacronym{mimo}{MIMO}{Multiple Input Multiple Output}
\newacronym{mmwave}{mmWave}{millimeter wave}
\newacronym{mptcp}{MPTCP}{Multipath TCP}
\newacronym{mr}{MR}{Maximum Rate}
\newacronym{mss}{MSS}{Maximum Segment Size}
\newacronym{mtd}{MTD}{Machine-Type Device}
\newacronym{mtu}{MTU}{Maximum Transmission Unit}
\newacronym{nfv}{NFV}{Network Function Virtualization}
\newacronym{nlos}{NLOS}{Non Line of Sight}
\newacronym{nlosb}{NLOSb}{Building Non Line of Sight}
\newacronym{nlosv}{NLOSv}{Vehicle Non Line of Sight}
\newacronym{nr}{NR}{New Radio}
\newacronym{ofdm}{OFDM}{Orthogonal Frequency Division Multiplexing}
\newacronym{pdcch}{PDCCH}{Physical Downlonk Control Channel}
\newacronym{pdcp}{PDCP}{Packet Data Convergence Protocol}
\newacronym{pdsch}{PDSCH}{Physical Downlink Shared Channel}
\newacronym{pdu}{PDU}{Packet Data Unit}
\newacronym{pf}{PF}{Proportional Fair}
\newacronym{pgw}{PGW}{Packet Gateway}
\newacronym{phy}{PHY}{Physical}
\newacronym{pbch}{PBCH}{Physical Broadcast Channel}
\newacronym[plural=\gls{mme}s,firstplural=Mobility Management Entities (MMEs)]{mme}{MME}{Mobility Management Entity}
\newacronym{prb}{PRB}{Physical Resource Block}
\newacronym{pss}{PSS}{Primary Synchronization Signal}
\newacronym{pucch}{PUCCH}{Physical Uplink Control Channel}
\newacronym{pusch}{PUSCH}{Physical Uplink Shared Channel}
\newacronym{rach}{RACH}{Random Access Channel}
\newacronym{ran}{RAN}{Radio Access Network}
\newacronym{red}{RED}{Random Early Detection}
\newacronym{rf}{RF}{Radio Frequency}
\newacronym{rlc}{RLC}{Radio Link Control}
\newacronym{rlf}{RLF}{Radio Link Failure}
\newacronym{rrc}{RRC}{Radio Resource Control}
\newacronym{rrm}{RRM}{Radio Resource Management}
\newacronym{rr}{RR}{Round Robin}
\newacronym{rs}{RS}{Remote Server}
\newacronym{rsrp}{RSRP}{Reference Signal Received Power}
\newacronym{rss}{RSS}{Received Signal Strength}
\newacronym{rtt}{RTT}{Round Trip Time}
\newacronym{rw}{RW}{Receive Window}
\newacronym{rx}{RX}{Receiver}
\newacronym{sa}{SA}{standalone}
\newacronym{sack}{SACK}{Selective Acknowledgment}
\newacronym{sap}{SAP}{Service Access Point}
\newacronym{sch}{SCH}{Secondary Cell Handover}
\newacronym{scoot}{SCOOT}{Split Cycle Offset Optimization Technique}
\newacronym{sdma}{SDMA}{Spatial Division Multiple Access}
\newacronym{sinr}{SINR}{Signal to Interference plus Noise Ratio}
\newacronym{sm}{SM}{Saturation Mode}
\newacronym{snr}{SNR}{Signal to Noise Ratio}
\newacronym{son}{SON}{Self-Organizing Network}
\newacronym{ss}{SS}{Synchronization Signal}
\newacronym{srs}{SRS}{Sounding Reference Signal}
\newacronym{sss}{SSS}{Secondary Synchronization Signal}
\newacronym{tb}{TB}{Transport Block}
\newacronym{tcp}{TCP}{Transmission Control Protocol}
\newacronym{tdd}{TDD}{Time Division Duplexing}
\newacronym{tdma}{TDMA}{Time Division Multiple Access}
\newacronym{tfl}{TfL}{Transport for London}
\newacronym{tm}{TM}{Transparent Mode}
\newacronym{prr}{PRR}{Packet Reception Ratio}
\newacronym{trp}{TRP}{Transmitter Receiver Pair}
\newacronym{tti}{TTI}{Transmission Time Interval}
\newacronym{ttt}{TTT}{Time-to-Trigger}
\newacronym{tx}{TX}{Transmitter}
\newacronym{ue}{UE}{User Equipment}
\newacronym{ul}{UL}{Uplink}
\newacronym{uml}{UML}{Unified Modeling Language}
\newacronym{um}{UM}{Unacknowledged Mode}
\newacronym{utc}{UTC}{Urban Traffic Control}
\newacronym{vm}{VM}{Virtual Machine}
\newacronym{rsrq}{RSRQ}{Reference Signal Received Quality}
\newacronym{rssi}{RSSI}{Received Signal Strength Indicator}
\newacronym{crs}{CRS}{Cell Reference Signal}
\newacronym{v2v}{V2V}{Vehicle-to-Vehicle}
\newacronym{v2i}{V2I}{Vehicle-to-Infrastructure}
\newacronym{v2n}{V2N}{Vehicle-to-Network}
\newacronym{v2x}{V2X}{Vehicle-to-Everything}
\newacronym{vn}{VN}{Vehicular Node}
\newacronym{dsrc}{DSRC}{Dedicated Short Range Communication}
\newacronym{ci}{CI}{context information}
\newacronym{voi}{VoI}{value of information}
\newacronym{gps}{GPS}{Global Positioning System}
\newacronym{qos}{QoS}{Quality of Service}
\newacronym{ml}{ML}{Machine Learning}
\newacronym{ahp}{AHP}{Analytic Hierarchy Process}
\newacronym{lidar}{LIDAR}{Light Detection and Ranging}
\newacronym{sumo}{SUMO}{Simulation of Urban MObility}
\newacronym{wave}{WAVE}{Wireless Access in Vehicular Environment}
\newacronym{c-its}{C-ITS}{Connected-Intelligent Transportation System}
\usepackage{cite}
\usepackage{bbold}
\usepackage{bbm}

\usepackage{array}
\newcolumntype{?}{!{\vrule width 1.5pt}}
\usepackage{makecell}
\usepackage{mdframed}
\usepackage[many]{tcolorbox}
\usepackage{empheq}

\newcommand*\justify{%
  \fontdimen2\font=0.4em
  \fontdimen3\font=0.2em
  \fontdimen4\font=0.1em
  \fontdimen7\font=0.1em
  \hyphenchar\font=`\-
}

\newtcbox{\mybox}[1][]{nobeforeafter,math upper,tcbox raise base,
  enhanced,frame hidden,boxrule=0pt,interior style={top color=green!10!white,
  bottom color=green!10!white,middle color=green!50!yellow},
  fuzzy halo=1pt with green,drop large lifted shadow,#1}

\usepackage{siunitx}
\usetikzlibrary{fadings}

\tikzfading[name=middle,
            top color=transparent!100,
            bottom color=transparent!100,
            middle color=transparent!20]

\usetikzlibrary{arrows,automata,calc,shapes, positioning,shadows,shadows.blur,shapes.geometric}

\def \eqPLLTE{
\begin{equation}
P_{\rm LOS}^{\rm UMi}(d) = \min \Big( \tfrac{0.018}{d},1 \Big) \left[ 1-\exp\left(\frac{-d}{0.063}\right)\right]+
\exp\left(\frac{-d}{0.063}\right),
\end{equation}
}

\def \eqPLmmwave{
 \begin{empheq}[left = \empheqlbrace]{align}
&P_{\rm LOS}^{\rm RMa}(d) = \exp\left( -\frac{d-10}{1000}\right) \text{ for } 10 \text{ m} < d\;;\\
&P_{\rm LOS}^{\rm UMi}(d) = \tfrac{18}{d}+\exp\left( -\frac{d}{36}\right)\left( 1-\frac{18}{d}\right) \text{ for } 18 \text{ m} < d\;.
\end{empheq}
}

\begin{document}
\pagenumbering{gobble}


\title{\fontsize{23}{29}\selectfont LTE and Millimeter Waves for  V2I  Communications: an End-to-End Performance Comparison}

\author{\IEEEauthorblockN{Marco Giordani, Andrea Zanella, Michele Zorzi}
\IEEEauthorblockA{\\
Department of Information Engineering, University of Padova, Italy\\
 Email:{\{giordani, zanella, zorzi\}@dei.unipd.it \vspace{-0.33cm}}}}

\maketitle

\begin{abstract}
The Long Term Evolution (LTE) standard enables, besides cellular connectivity, basic automotive services to promote road safety through vehicle-to-infrastructure (V2I) communications. 
Nevertheless, stakeholders and research institutions, driven by the ambitious technological advances expected from fully autonomous and intelligent transportation systems,  have recently investigated new radio technologies as a~means to support vehicular applications. 
In particular, the millimeter wave (mmWave) spectrum holds great promise 
because of the large available bandwidth that may provide the required link capacity.
Communications at high frequencies, however, suffer from severe propagation and absorption loss, which may cause communication disconnections especially considering high mobility scenarios.
It is therefore important to validate, through simulations,  the actual feasibility of establishing V2I communications in the above-6 GHz bands.
Following this rationale, in this paper we provide the first comparative end-to-end evaluation of the performance of the LTE and mmWave technologies in a vehicular scenario. The simulation framework includes detailed measurement-based channel models as well as the full details of MAC, RLC and transport protocols. 
Our results show that, although LTE  still represents a promising access solution to guarantee robust and~fair connections, mmWaves satisfy the foreseen extreme throughput demands of most emerging automotive~applications.

\end{abstract}


\section{Introduction} 
\label{sec:introduction}

\begin{figure*}[t!]
 \centering
 \captionof{table}{\footnotesize  Performance requirements of different use cases specified by the 3GPP in~\cite{3GPP_22186}.}
 \label{tab:requirements}
 \includegraphics[width = 0.88\textwidth]{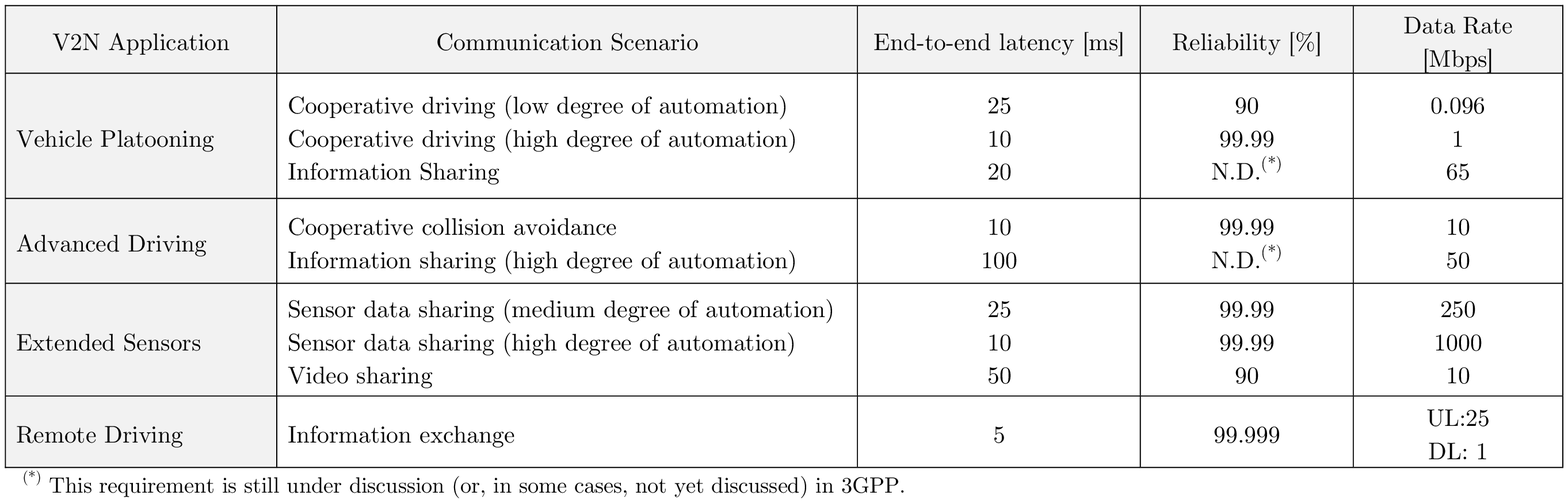}
    \label{fig:requirements}
 \end{figure*}

In recent years, the automotive industry has rapidly evolved towards the development of advanced automotive services offering safer traveling, improved traffic management, and support to infotainment applications.
A key enabler of this evolution is \gls{v2i} communication, which allows vehicles to communicate with road-side infrastructures and the Internet. 
The \gls{lte} standard presently represents the principal wireless interface offering \gls{v2i} transmission services~\cite{araniti2013LTE}.
However, future \glspl{c-its} will need to satisfy unprecedentedly stringent demands in terms of latency and throughput (i.e., in the order of terabytes per driving hour according to some estimates~\cite{lu2014connected}) which may saturate the capacity of traditional technologies for vehicular communications.
In this perspective, industry players have devoted efforts into specifying new communication solutions as enablers of the performance requirements of next-generation automotive networks. The \gls{mmwave} spectrum -- roughly above 10 GHz~\cite{magazine2016_Heath} -- currently holds great promise because of the large available bandwidth that may  guarantee data rates in the order of multi-gigabit-per-second.

Although the application of  \glspl{mmwave} in a vehicular context is not new (automotive radars operating in the 77 GHz band are already in the market~\cite{hasch2012millimeter}),  the severe isotropic path loss and blockage absorption of \gls{mmwave} signals, as well as the increased Doppler effect in high mobility scenarios,  make the design of wireless systems in the above-6 GHz bands very challenging~\cite{MOCAST_2017}.
Before unleashing the potential of this technology into a \gls{v2i} deployment, it is therefore fundamental to  validate the practical feasibility of designing 	\gls{mmwave}-aware strategies in view of the strict requirements and the specific features of future transportation~systems.

Motivated by the above introduction, our paper targets the following objectives.
First, we provide the first comprehensive end-to-end performance evaluation of the \gls{mmwave} and the \gls{lte} paradigms in a vehicular environment.  
In particular, we characterize the system throughput and latency when varying the density of network infrastructures, the target application's demands  and the channel model.
We also consider both urban and highway scenarios, to characterize different mobility and propagation regimes.
Unlike traditional performance analyses, e.g., \cite{mase2008performance,performance2018giordani,beamDesignV2I_Heath}, which rely on \gls{phy} or \gls{mac} layer quality metrics (e.g., achievable transmission range or packet transmission probability), our work investigates the impact of the upper layers on the network behavior, thereby guaranteeing more accurate system-level analyses. 
Moreover, unlike analytical evaluations, e.g.~\cite{giordani2018coverage,Tassi17_Highway}, which typically adopt conservative assumptions on the signal propagation, our paper considers full-stack simulations, which allow to estimate the system performance accounting for  detailed protocol implementations. 
Second, we evaluate through numerical comparisons whether the \gls{mmwave} technology might represent a more promising solution in creating a safer and more efficient driving ecosystem than its \gls{lte} counterpart.
Third, based on our extensive simulation results, we provide guidelines to identify the most appropriate network interface that satisfies \gls{v2i} service requirements while establishing high-capacity channels, a research task that, to date, has not been throughly investigated yet.

Our simulation campaign proves that, although \gls{lte} still represents a promising access solution to maintain robust communications,  mmWave technology emerges as an enabler of the foreseen extreme throughput demands of future automotive applications.
Moreover, we demonstrate that, while densification is beneficial to urban \gls{mmwave} deployments  for both throughput and latency, it may have a  negative impact on the performance of  \gls{lte} systems and in highway scenarios.
Our study reveals also several important findings on the interaction between the transport layer mechanisms and the underlying physical~propagation. 

The rest of the paper is organized as follows. Sec.~\ref{sec:v2n_radio_technologies_an_overview} overviews the characteristics of the \gls{lte} and the \gls{mmwave} radios as enabling technologies for \gls{v2i} communications.
Sec.~\ref{sec:evaluation_scenarios_and_methodology} describes our simulation setup and Sec.~\ref{sec:end_to_end_evaluation} presents our main findings and comparative results. Finally,  Sec.~\ref{sec:conclusions} concludes the paper and discusses possible research~extensions.

\section{V2I Radio Technologies: An Overview} 
\label{sec:v2n_radio_technologies_an_overview}

\glspl{cav}, when fully commercialized, will address the societal and business trends of 2020 and beyond, and will have ever more stringent regulations in terms of road safety and traffic efficiency~\cite{boban2018connected}.
In this regard, the \gls{3gpp}, in its Release 15, defines new use cases specific to future vehicular services whose requirements, although not yet fully specified, have already been outlined in~\cite{3GPP_22186}, as summarized in Table~\ref{tab:requirements}.

\begin{itemize}
	\item \emph{Vehicles Platooning.} It refers to the set of services that enable the vehicles to  cooperatively travel in close proximity to one another at highway speeds. The data rate ranges from a few Kbps up to 65 Mbps depending on whether sensor sharing is required, while the latency ranges from 10 ms to 500 ms depending on the inter-vehicle distance. Vehicle platooning poses also very strict requirements in terms of connection reliability.

	\item \emph{Advanced Driving.} It enables semi- or fully-automated driving through persistent dissemination of perception data.
	While the required data rate is relatively low (i.e., less than 50 Mbps), the latency must be very small (i.e., less than 100 ms for high degree of automation) to ensure prompt reactions to unpredictable events.

	\item \emph{Extended Sensors.} It enables the exchange of raw or processed data gathered through local sensors, thereby enhancing the perception range of the vehicles beyond the capabilities of their on-board instrumentation. 
	The data rate demands are proportional to the resolution of the acquired sensory data and range from around 10 Mbps for a 300-beam 32-bit LIDAR  up to approximately 1 Gbps for high-quality uncompressed camera images~\cite{kim2015multivehicle}.  
	Due to the sensitive nature of the exchanged information, the maximum tolerable latency  varies from approximately 3 ms up to 100 ms for lower degrees of automation.

	\item \emph{Remote Driving.} It enables remote control of a vehicle by either a human operator or cloud computing, to support coordination between vehicles in dangerous conditions. Remote driving services require high uplink throughput connections (i.e., around 25 Mbps) and an end-to-end latency lower than 5 ms for fast vehicle teleoperations. Moreover, ultra-high reliability (i.e., 99.999\% or higher) shall be guaranteed to avoid application malfunctions.

\end{itemize}
 
Given the variety of automotive services and the heterogeneity of their requirements, it is unlikely that \gls{v2i} communications will be supported by a single radio  solution, rather the orchestration of multiple technologies is recommended. In this section we therefore overview the characteristics of candidate radio interfaces currently being considered for  \gls{v2i} communications, i.e., the LTE and the mmWave technologies.

\subsection{Long Term Evolution (LTE)} 
\label{sub:long_term_evolution_}

Since its inception, the \gls{lte} cellular technology, operating in the sub-6 GHz spectrum, has represented an ideal candidate to support \gls{v2i} operations~\cite{3GPP_22885}.
First, LTE relies on a capillary deployment of \glspl{enb} offering wide area coverage and long-lived connectivity.
Second, resource allocation is centrally managed by an \gls{enb} at every transmission opportunity, thereby satisfying service quality constraints while managing priorities in case of \gls{v2i} applications competing for resources~\cite{boban2018connected}.
Third, LTE operates through omnidirectional transmissions and therefore supports  broadcast data distribution~\cite{araniti2013LTE}.
Fourth, the LTE interface may guarantee transfer latencies in the radio access theoretically lower than 100 ms,  which is particularly beneficial for delay-sensitive vehicular~applications.

Nevertheless, LTE was originally designed for mobile broadband traffic and its capability to support \gls{v2i} communications is still an open issue.
The main concern comes from LTE’s  architecture, that is configured to keep non-active terminals in idle mode: transitions to connected mode  may require several seconds~\cite{36600}, which is intolerable for vehicular services. 
The access and transmission latency also increases with the number of users in the cell, thus raising issues.
Moreover, despite the almost ubiquitous coverage of LTE, still the connection may not be always available (e.g., in underground areas).
Finally, LTE offers limited downlink~capacity, which might not be enough to satisfy the  requirements of some  \gls{v2i}~applications.

\subsection{Millimeter Waves (mmWaves)} 
\label{sub:millimeter_waves_}

Recently, the \gls{mmwave} band has been investigated as a means to enhance automated driving and address the stringent throughput and latency demands of emerging vehicular applications.
These frequencies, combined with high-order modulation and  \gls{mimo} techniques, offer orders of magnitude higher bitrates than legacy vehicular technologies~\cite{magazine2016_Heath}.
Moreover, unlike in  LTE, \gls{mmwave} systems operate through highly directional communications which tend to isolate the users and deliver reduced interference. Inherent security and privacy is also improved because of the short-range transmissions which are typically established~\cite{rappaport2014millimeter}.

Although \gls{mmwave}-assisted \gls{v2i} operations are very attractive from the throughput perspective, they still pose significant challenges~\cite{pi2011introduction,MOCAST_2017}.
Signals propagating in the \gls{mmwave} spectrum suffer from severe path loss and susceptibility to shadowing, thereby preventing long-range transmissions (assuming isotropic propagation).
Furthermore, directionality requires precise beam alignment of the transmitter and the receiver.
In high density and/or high mobility scenarios, the corresponding peer may change frequently,
thus implying increased control overhead and communication disconnections.
Additionally,  \gls{mmwave} links are highly sensitive to blockage
and have ever more stringent requirements on electronic components, size, and power consumption.
Given that the challenging radio conditions derived from the mobility of vehicles are further exacerbated considering the dynamic topology of the  vehicular networks, 
the direct applicability of  \gls{mmwave} technology to a \gls{v2i} deployment is still not clear and has become a research focus in the area of intelligent autonomous~systems~\cite{3GPP_181435}.

\section{Evaluation Methodology and Setup} 
\label{sec:evaluation_scenarios_and_methodology}

In this section we give an overview of the methodology we use to assess the performance of the \gls{v2i} deployment. 
In detail, in Sec.~\ref{sub:ns_3_integration} we briefly describe the architecture of the LTE and the mmWave modules in ns-3, while in Secs.~\ref{sub:simulation_setup} and \ref{sub:performance_metrics} we introduce our system-level simulation parameters and performance metrics, respectively.

\subsection{The ns-3 Architecture} 
\label{sub:ns_3_integration}

Our performance evaluation is conducted using ns-3~\cite{henderson2008network},  an open source software which allows the simulation of complex networks  with a very high level of detail. The ns-3 simulator features both an LTE and a mmWave full protocol stack, as described in the following paragraphs.\\

\textbf{LTE Module.}
The LTE ns-3 module, designed by Centre Tecnològic Telecomunicacions Catalunya (CTTC) in 2011, provides a basic implementation of LTE devices, including propagation models, \gls{phy} and \gls{mac} layers.
A complete description of the LTE module is presented in~\cite{piro2011lte}: it features (i) a basic implementation of both the \gls{ue} and the \gls{enb} devices, (ii) \gls{rrc} entities for both the \gls{ue} and the \gls{enb}, (iii)  handover mechanisms for \gls{ue} mobility management, (iv) \gls{rrm} of the data radio bearers, the MAC queues and the \gls{rlc} instances,  (v) support for both uplink and downlik packet scheduling, and (vi) a PHY layer model with \gls{rb} level granularity. 

The path loss  is based on pure geometric considerations which deterministically evaluate whether the \gls{v2i} link is blocked or not by buildings.
In this paper, we further extend the LTE module introducing a probabilistic model for the characterization of the channel between  the \gls{ue} and the \gls{enb} devices as a function of the distance $d$ for both \gls{los} and \gls{nlos} propagation~\cite{3GPP_LTE_channel}. In case of urban (UMi) scenario, a vehicle is in \gls{los} with probability
\medmuskip=0mu
\thickmuskip=0mu
\eqPLLTE
\medmuskip=6mu
\thickmuskip=6mu
and in \gls{nlos} with  probability $P_{\rm NLOS}(d) = 1-P_{\rm LOS}^{\rm UMi}(d)$. 
The path loss, for both \gls{los} and \gls{nlos} cases, is implemented in the new \texttt{\justify Lte3gppPropagationLossModel} class following the model in~\cite{3GPP_LTE_channel}.
In case of highway (RMa) scenario, the channel characterization follows the Friis free-space model.
In addition, we consider a fast Rayleigh fading, which is modeled as a stochastic gain with unit power (in linear scale).\\

\textbf{mmWave Module.}
The \gls{mmwave} ns-3 module, designed by NYU and the University of Padova in 2015, builds upon the LTE module and implements a complete 3GPP-like cellular protocol stack.
A complete description of the \gls{mmwave} module is presented in~\cite{mezzavilla2018end}: it features (i) a custom PHY/MAC layer implementation for both \gls{ue} and \gls{enb} devices, (ii) support for directional transmissions through analog beamforming, (iii)  a dynamic Time Division Duplexing (TDD) scheme designed for low latency communications, (iv) \gls{rlc}, \gls{pdcp} and \gls{rrc} layers, and (v) a complete TCP/IP protocol suite.

The propagation model is based on the 3GPP channel model for frequencies above 6 GHz~\cite{38901}, which characterizes the time correlation among the channel impulse responses  to account for spatial consistency. The \gls{los} probability for both UMi and RMa scenarios, in case the propagation is free of obstructions, is given~by
\medmuskip=0mu
\thickmuskip=0mu
\eqPLmmwave
\medmuskip=6mu
\thickmuskip=6mu

The path loss, for both \gls{los} and \gls{nlos} cases, is finally implemented in the \texttt{MmWave3gpp}\allowbreak\texttt{PropagationLossModel} class, as described in~\cite[Sec. 7.4]{38901}.
Moreover, since the effects of high mobility result in rapidly time-varying channels at \glspl{mmwave}, ns-3 implements a very detailed fading model in the \texttt{MmWave3gppChannel} class.
In particular, the model characterizes spatial clusters, subpaths, angular beamspreads and the Doppler shift, which is a function of the total angular dispersion, carrier frequency and mobile velocity.

       


\renewcommand{\arraystretch}{0.9}
\begin{table}[t!]
\small
\centering
\caption{Main simulation parameters.}
\begin{tabularx}{0.9\columnwidth}{ @{\extracolsep{\fill}} lll}
\toprule
Parameter & Value  \\ \midrule
  mmWave bandwidth $W_{\rm mmW}$ &  $1$ GHz & \\
mmWave carrier frequency $f_{\rm c, mmW}$ & $28$ GHz \\
eNB  array size  $M_{\rm UPA,eNB}$ & $8\times8$  \\
Vehicle  array size  $M_{\rm UPA,V}$ & $4\times4$  \\
 LTE carrier frequency $f_{\rm c, LTE}$ & $2$ GHz \\
 LTE bandwidth $W_{\rm LTE}$ & $20$ MHz\\
 eNB density $\lambda_{\rm eNB}$ &\{4\dots80\}/km$^2$ \\
 TX power    $P_{\rm TX}$ & $30$ dBm \\
 Packet size $D$ & 1400 B \\
 Noise figure  NF & $5$ dB  \\
 Application rate $R$ & $\{224,11,1\}$ Mbps \\
 RLC buffer size $B_{\rm RLC}$ & 10 MB\\
 Vehicles per eNB$M_{V}$ & 10 \\
 RLC AM reordering timer  $\tau_{\rm RLC}$ &1 ms\\
 Number of simulation runs $N_{\rm sim}$ & 100 \\
\bottomrule
\end{tabularx}
\label{tab:params}
\end{table}



\subsection{Simulation Setup} 
\label{sub:simulation_setup}

The simulation parameters are based on realistic system design considerations and are summarized in Table \ref{tab:params}.
The \gls{mmwave} and \gls{lte} \glspl{enb} are deployed over an area of $500\times500$ meters
according to a  Poisson Point Processes (PPP) of density $\lambda_{\rm eNB}$, with $\lambda_{\rm eNB}$ varying from 4 to 80 eNB/km$^2$ (the trade-off involves signal coverage and deployment cost).
We also deploy an average of $M_{\rm V}=10$ vehicles per eNB, as foreseen in~\cite{3GPP_38913} for a dense environment.
We consider urban  and highway scenarios, i.e., UMi-Street-Canyon and RMa, according to the \gls{3gpp} terminology, to characterize different mobility and propagation regimes, as described in Sec.~\ref{sub:ns_3_integration}. 

At the \gls{phy} layer, LTE eNBs operate in the 2 GHz  band, with 20 MHz of bandwidth and omnidirectional transmissions. 
Conversely, \gls{mmwave} eNBs operate at 28 GHz with 1 GHz of bandwidth and are equipped with \glspl{upa} of $8\times 8$ elements to establish directional communications through beamforming. Vehicles are also equipped with  $4\times 4$ \glspl{upa}.\footnote{For a discussion on the impact of the antenna array size on the overall system performance, we refer the interested reader to our previous work ~\cite{performance2018giordani}.} For both \gls{lte} and \gls{mmwave} systems, the transmission power and noise figure are set to $P_{\rm TX} = 30$ dBm and NF $=5$ dB, respectively.

The \gls{mac} layer performs \gls{harq} to enable fast retransmissions in case of corrupted~receptions, and the \gls{rlc} layer, whose buffer is $B_{\rm RLC} = 10$ MB, uses \gls{am} to offer additional~reliability.

\gls{udp} is used at the transport layer. Each vehicular application generates packets of $D = 1400$ bytes at a constant interarrival rate ranging from $\tau_{\rm min}= 50$ $\mu s$  $\tau_{\rm max}= 10000$ $\mu s$, corresponding to application rates ranging from $R_{\rm max} \simeq 224$ Mbps to $R_{\rm min} \simeq 1$ Mbps, to test the performance of \gls{lte} and \glspl{mmwave} in relation with different service requirements.
According to Table~\ref{tab:requirements}, high-rate transmissions are compatible with \gls{v2i} applications offering extended sensor sharing services, while lower source rates are typical of platooning systems having very stringent requirements in terms of communication delay and reliability  but for which the size of the exchanged messages is reasonably small.

\begin{figure}[t!]
\centering
    \setlength{\belowcaptionskip}{-0.53cm}
    \setlength\fwidth{0.78\columnwidth}
    \setlength\fheight{0.55\columnwidth}
    \includegraphics[width = 0.9\columnwidth]{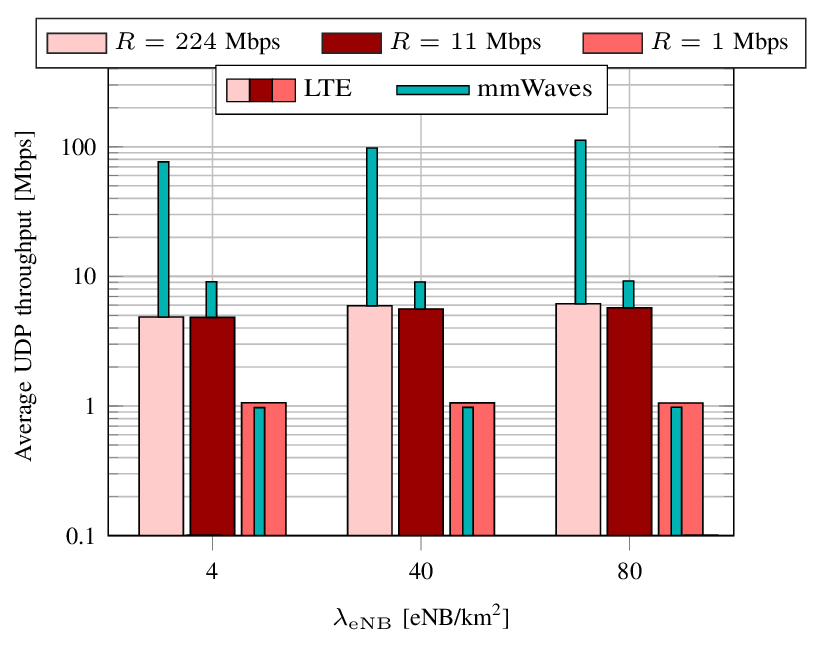}
    \caption{Average UDP throughput vs. $\lambda_{\rm eNB}$ and the application rate in urban scenarios. Narrow (wide) bars refer to a mmWave (LTE) system.}
    \label{fig:avgThr}
\end{figure}



\subsection{Performance Metrics} 
\label{sub:performance_metrics}

The statistical results are derived through a Monte Carlo approach, where 100 independent simulations are repeated to get different quantities of interest. 
In particular, we analyze the following end-to-end performance metrics.

\begin{itemize}
  \item  \textit{Average UDP throughput,}  which is computed as the total number of received bytes divided by the total simulation time, averaged over the $N_{\rm sim}$ simulations.
  \item \textit{Total UDP throughput}, the sum of the throughput~experienced by all vehicles within the coverage  of a given~eNB.
  \item \textit{5th (and 10th) percentile UDP throughput}, the average throughput relative to the worst 5\% (10\%) of the vehicles (which represents the performance of cell-edge nodes,  the most resource-constrained network entities).
  \item \textit{Average UDP latency}, which is measured for each packet, from the time it is generated at the application layer to when it is successfully received assuming perfect beam alignment (it is therefore the latency of only the correctly received packets).
  \item \textit{Jain's fairness index}, which is used to determine whether vehicles are receiving a fair share of the cell resources. This index is defined as
  \begin{equation}
J = \frac{\left(\sum_{i=1}^{M_V} S_i\right)^2}{{M_V} \sum_{i=1}^{M_V} S_i^2},
\label{eq:Jain}
\end{equation}
where $M_V$ is the number of users in the cell and $S_i$ is the throughput experienced by the i-th vehicle. The result ranges from $1/M_V$ (most unfair) to 1 (most fair).
\end{itemize}

\section{End-to-End Performance Evaluation} 
\label{sec:end_to_end_evaluation}

In this section we provide some numerical results to evaluate the end-to-end performance of the LTE and the mmWave technologies in a V2I scenario.\\

\textbf{UDP Throughput.}
Fig.~\ref{fig:avgThr} shows the average experienced UDP throughput for different eNB densities.
We observe that, for the low source rate scenario (i.e., $R = 1$ Mbps),  both LTE and mmWave systems  deliver comparable values of throughput,  which is almost equal to the full  rate offered by the application layer. 
Conversely, higher-rate applications  (i.e., $R=11$ Mbps and $R=224$ Mbps) are not well-supported by LTE connections  which are constrained by the limited capacity of the low-bandwidth physical channel.
The mmWave spectrum, in turn, offers orders of magnitude higher data rates than  lower frequencies even in case of congested channels (110 Mbps vs. 7 Mbps  for $\lambda_{eNB}=80$ eNB/km$^2$), although still not satisfying the requirements of the most demanding vehicular~services.

Moreover, we see that the throughput generally increases with the eNB density, as a consequence of stronger channels.
The effect of densification is particularly evident  for mmWave networks (i.e., the throughput increases by more than 50\% from 4 to 80 eNB/km$^2$ for $R=224$ Mbps) since the endpoints are progressively closer thus guaranteeing improved signal quality and higher received power.
On the other hand, densification delivers very negligible improvements for the LTE case due to the more serious impact of interference in case of omnidirectional communications.
The above discussion exemplifies how, unlike in legacy networks, the harsh propagation characteristics of the above-6 GHz bands advocate  for a high-density deployment of eNBs, to guarantee   \gls{los}  at any given time and decrease the outage probability~\cite{RanRapE:14}.\footnote{Overdensification, in turn, might lead to performance degradation if the number of handovers increases uncontrollably, e.g.,  in high mobility scenarios.}
We finally highlight that, for low-rate applications,  the UDP traffic injected in the system is sufficiently well handled by the buffer, with no overflow, also considering sparsely deployed~networks.
\\

\begin{figure}[t!]
\centering
    \setlength{\belowcaptionskip}{-0.53cm}
    \setlength\fwidth{0.73\columnwidth}
    \setlength\fheight{0.55\columnwidth}
    \includegraphics[width = 0.9\columnwidth]{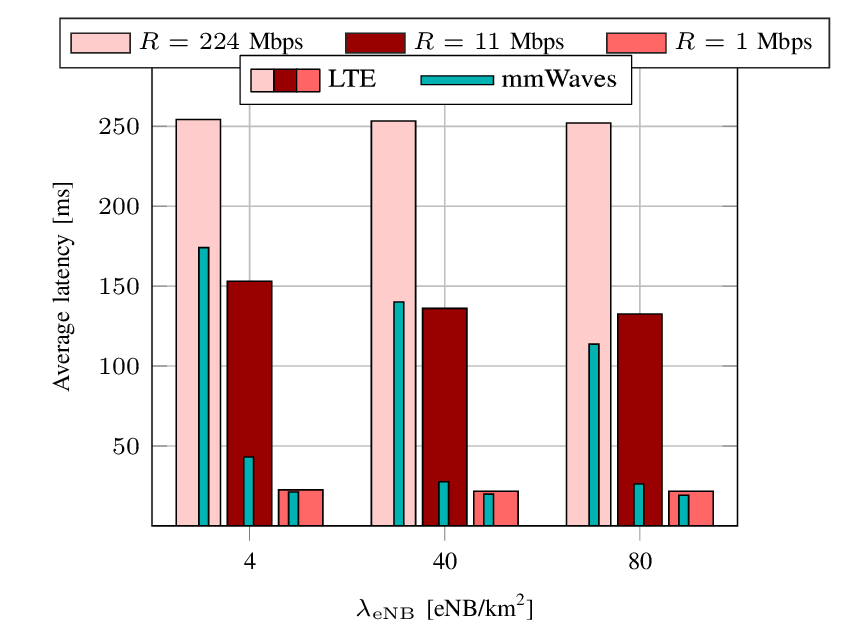}
    \caption{Average UDP latency vs. $\lambda_{\rm eNB}$ and the application rate in urban scenarios. Narrow (wide) bars refer to a mmWave (LTE)~system.}
    \label{fig:avglat}
\end{figure}

\textbf{UDP Latency.}
In Fig.~\ref{fig:avglat} we measure the average communication latency as a function of  $\lambda_{eNB}$ for different application rates.\footnote{The above results were derived considering an RLC buffer of 10 MB. However, the buffer size is critical for the performance of the network~\cite{zhang2019will}.}
We observe that, for $R = 1$ Mbps, both LTE and mmWave guarantee very low latency (i.e., below 20 ms) since the MAC buffers are empty most of the time. 
For $R = 11$ Mbps, although the two technologies were proven to offer comparable average throughput (7 Mbps vs. 9 Mbps, respectively,  for $\lambda_{eNB}=40$ eNB/km$^2$),  mmWave systems guarantee 5 times lower latency than legacy systems, which  cannot ensure time critical message dissemination in case of highly saturated channels.
For higher application rates, the end-to-end latency increases uncontrollably in all investigated configurations as a consequence of longer queueing at the MAC layer, although the overall average latency for the mmWave deployment (i.e., around 150 ms for $\lambda_{eNB}=40$ eNB/km$^2$) is still more than 50\%  lower than its LTE counterpart.

Additionally, Fig.~\ref{fig:avglat} illustrates that increasing the eNB density in  mmWave scenarios has beneficial effects in terms of latency reduction (more than 35\% as a results of densification from 4 to 80 eNB/km$^2$ for $R = 224$ Mbps) as compared to a reduction of only 1\% in case of LTE~connections.
According to Table~\ref{tab:requirements}, these latencies would likely satisfy the envisioned requirements for most next-generation V2I applications (e.g., those supporting advanced driving services).\\

\begin{figure}[t]
	\centering
	 \setlength{\belowcaptionskip}{-0.43cm}
	\includegraphics[trim={0 0.2cm 0 0},clip,width = 0.92\columnwidth]{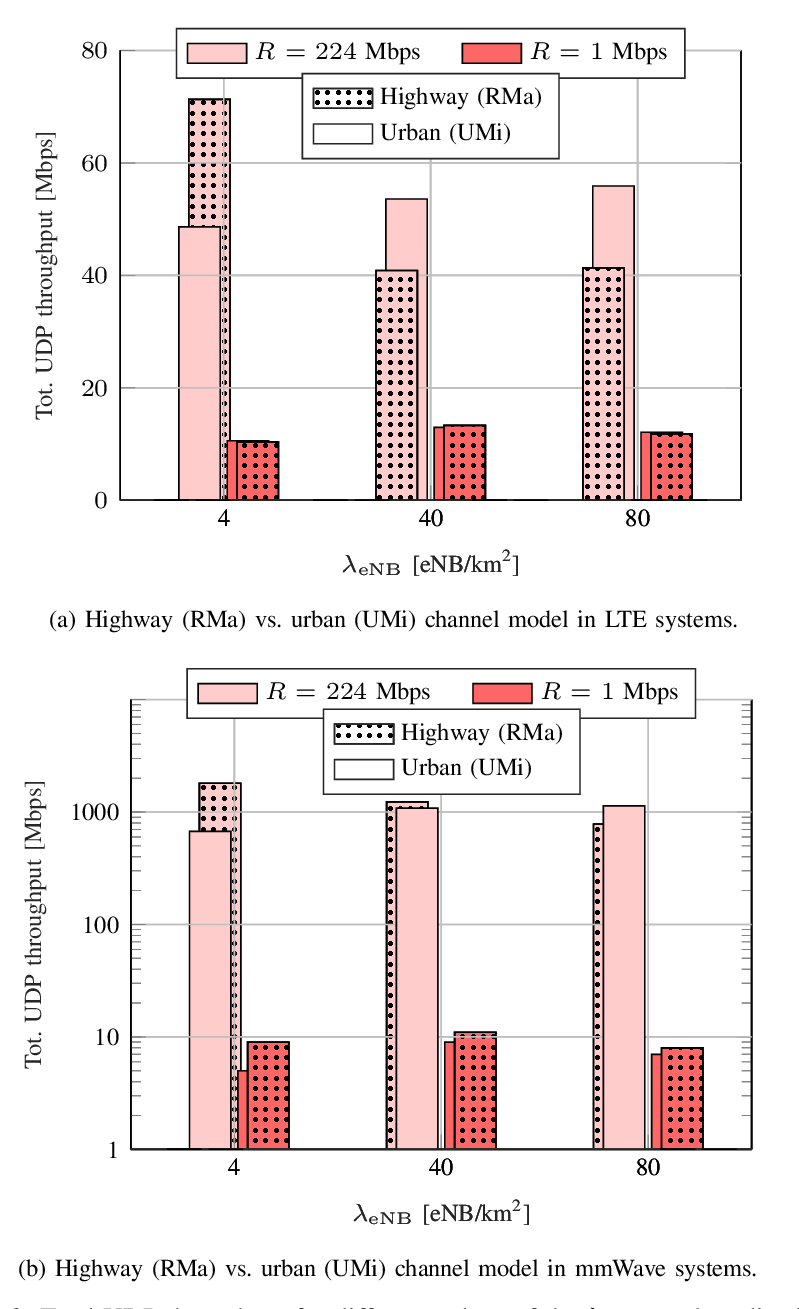}
  \vspace{0.1cm}
	\caption{Total UDP throughput  for different values of the $\lambda_{\rm eNB}$ and application rate. Dotted (straight) bars refer to a highway (urban) scenario. The performance of the LTE and the mmWave technologies are compared.}
	\label{fig:totThr} 
\end{figure}

\begin{figure}[t]
  \centering
  \setlength{\belowcaptionskip}{-0.63cm}
  \includegraphics[width = 0.95\columnwidth]{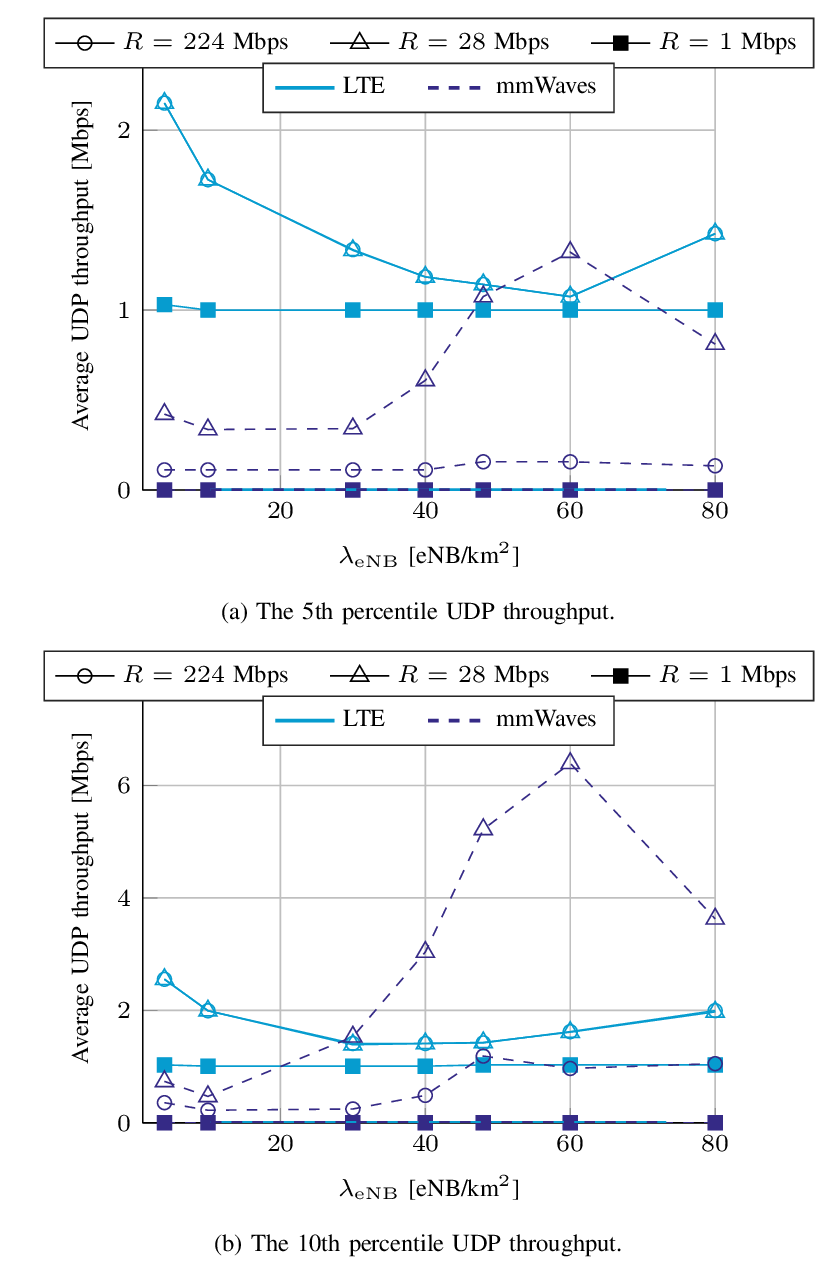}
  \caption{The 5th and 10th percentile UDP throughput vs. $\lambda_{\rm eNB}$ and the application rate in urban scenarios. The performance of the LTE and the mmWave technologies are compared. }
  \label{fig:percentile} 
\end{figure}

\textbf{UMi vs. RMa Propagation.} In Fig.~\ref{fig:totThr} we plot the  total UDP throughput as a function of the eNB density for both highway (RMa) and urban (UMi)  scenarios.
Considering highly saturated channels (i.e., $R=224$ Mbps), RMa  generally guarantees throughput improvements (i.e., +40\% and +63\% for LTE and mmWave systems, respectively) with respect to UMi in case of  sparse, thus noise-limited, networks (i.e., $\lambda_{\rm eNB}=4$ eNB/km$^2$): despite the increased Doppler effect of high mobility scenarios, free-space propagation indeed results in reduced outage probability.
On the other hand, when pushing the network into interference-limited regimes, thus for   dense and extremely dense deployments, the gain progressively reduces with $\lambda_{\rm eNB}$ because of the increasing impact of the interference from the surrounding cells.
For LTE deployments (Fig.~\ref{fig:totThr}a),  the total throughput starts decreasing for $\lambda_{\rm eNB} \geq 40$ eNB/km$^2$  as a result of RMa propagation which, while generally ensuring better signal quality, increases interfering signal strength unintentionally due to the transition of a large number of interference paths from \gls{nlos} to  \gls{los}~\cite{ding2016performance}. 
For mmWave deployments (Fig.~\ref{fig:totThr}b), RMa propagation induces more than 30\% throughput decrease for $\lambda_{\rm eNB} = 80$ eNB/km$^2$ compared to UMi propagation.
In fact,
while in the highway environment the propagating signals
attenuate over distance following the square power law, i.e.,
 Friis’ law, the waveguide effect resulting
from the more likely signal reflections and scattering in dense urban canyons results
in reduced attenuation.
Moreover, the presence of blockages in the UMi scenario may actually reduce the impact of the interference from neighboring eNBs when the obstructions block the interfering signals~\cite{kim2018millimeter}.  

Considering non-congested scenarios (i.e., $R=1$ Mbps) instead, Fig.~\ref{fig:totThr} proves that the experienced throughput becomes independent of the eNB density and the propagation environment since both UMi and RMa channels, regardless of their propagation characteristics, can support well the loose requirements typical of low source rate V2I applications.\\


\begin{figure}[t!]
\centering
  \setlength{\belowcaptionskip}{-0.53cm}
    \setlength\fwidth{0.73\columnwidth}
    \setlength\fheight{0.55\columnwidth}
    \includegraphics[width = 0.9\columnwidth]{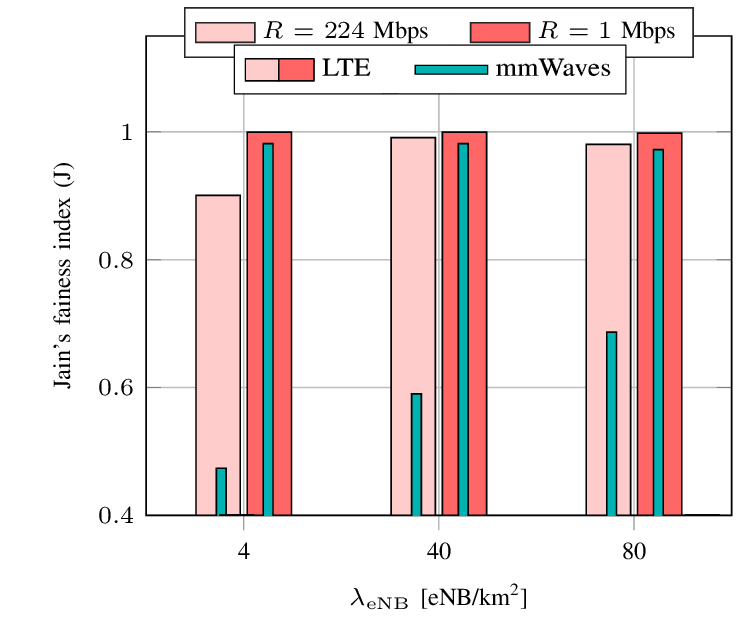}
    \caption{Jain's fairness index of the UDP throughput vs. $\lambda_{\rm eNB}$ and the  application rate in urban scenarios. Narrow (wide) bars refer to a mmWave (LTE) system.}
    \label{fig:jain}
\end{figure}

\textbf{5th/10th Percentile Throughput.}
Fig.~\ref{fig:percentile}  represents the 5th and 10th percentile  throughput  for different application rates.
First, we observe that, for sparsely deployed networks, LTE eNBs offer higher throughput to cell-edge vehicles than mmWave eNBs.
In this region,  most vehicles are in  \gls{nlos} and, unlike sub-6 GHz propagation, the challenging communication characteristics of high-frequency channels might result in outage to the serving cell.
Moreover, as edge vehicles are power-limited, they are unable to fully exploit the potential of the increased spectrum availability at \glspl{mmwave}~\cite{RanRapE:14}.  
Densification, in turn,  increases the \gls{los} probability and avoids the presence of coverage holes, thereby making the LTE and \gls{mmwave} radio solutions roughly comparable in terms of cell-edge throughput.

Second, Fig.~\ref{fig:percentile} shows that, for LTE deployments, the mutual interference from omnidirectional eNBs eventually impacts on the cell-edge  throughput, which decreases for increasing values of $\lambda_{\rm eNB}$.
Similarly, we see that, although the directional nature of \gls{mmwave} systems guarantees reduced interference,  there are some special cases where interference is not negligible, i.e., when $\lambda_{\rm eNB} > 45$ eNB/km$^2$ for $R=224$ Mbps. 

Third, while for LTE the 5th and 10th percentile rates reported in Fig.~\ref{fig:percentile}  compare similarly to the average values measured in Fig.~\ref{fig:avgThr},   mmWave systems alone cannot provide uniform capacity, with cell-edge users suffering significantly.
In particular,  the 5th percentile throughput experiences a dramatic 475 fold decrease (from around 100 Mbps to only 200 Kbps for $R=224$ Mbps and considering $\lambda_{\rm eNB} = 40$ eNB/km$^2$) with respect to average conditions, demonstrating a significant limitation of mmWaves under NLOS propagation.\\

\textbf{Fairness.} In Fig.~\ref{fig:jain} we plot  Jain’s fairness index, defined in Eq.~\eqref{eq:Jain}, for the average vehicle throughput considering both LTE and \gls{mmwave} scenarios.
Although fairness is not always required (e.g., some categories of applications, like those supporting time-critical safety operations, deserve prioritization), it still represents a major concern that should be taken into account to guarantee a minimum performance also to the cell-edge users (or, in general, to users experiencing bad channel conditions).
We observe that, for LTE systems,  Jain's index is very close to 1 for all density configurations,  indicating that (i) cell-edge vehicles experience a throughput comparable to that of other vehicles in the cell regardless of the source application rate, and (ii)  densification has a  negligible impact on the overall network performance.
Conversely, mmWave deployments are generally not  compatible with fairness.
In particular, the effect of a highly saturated network (i.e., $R=224$ Mbps) makes  Jain's index fall by an impressive 45\% (for $\lambda_{\rm eNB}=40$ eNB/km$^2$) compared to LTE propagation, as a result of the increased time-variability of the mmWave channel due to scattering and reflection from nearby obstructions, and due to higher Doppler spread.
However, such effect is partially mitigated considering denser deployments, i.e., as the probability of path loss outage decreases: in this case, the system is able to increase the coverage of cell-edge users, i.e., the most resource-constrained network entities, and consequently, provide more uniform quality of service throughout the network (for example, $J$ increases by more than 30\% when going from 4 to 80~eNB/km$^2$).





\section{Conclusions and Open Challenges} 
\label{sec:conclusions}

In this paper we provide the first end-to-end performance comparison between the LTE and the mmWave technologies in a \gls{v2i} deployment.
The impact of several automotive-specific parameters (i.e., the eNB density, the vehicular scenario and the application data rate) was investigated in terms of experienced throughput, communication latency and fairness.
We concluded that, although LTE delivers a good compromise between fairness and low latency, the combination of massive bandwidth and spatial degrees of freedom has the potential for \gls{mmwave} systems to meet some of the boldest requirements of next-generation transportation systems, including high peak per user data rate and very low latency, both in urban and high-mobility highway scenarios. 
We also demonstrated that, unlike in legacy \gls{v2i} networks, densification of mmWave eNBs is beneficial, for urban propagation, to decrease the outage probability and deliver uniform service quality throughout the cell.
In this context, the end-to-end communication performance can be improved by using multiple radios in parallel (i.e., \emph{hybrid networking}), to complement the limitations of each type of network and deliver more flexible and resilient~transmissions.

This work opens up interesting research directions.
In particular, we will consider more realistic traffic models and more complex evaluation scenarios to address dynamic topologies.
Moreover, we will design methods to  identify the best radio solution as a function of  channel characteristics and the environment in which the vehicles are~deployed.

\bibliographystyle{IEEEtran}
\bibliography{bibliography.bib}

\end{document}